\documentclass[12pt]{article}
\usepackage{latexsym}
\usepackage{amstext}
\usepackage{amsmath}
\usepackage[normalem]{ulem}
\usepackage{subfigure}
\usepackage{epsfig, graphicx}
\usepackage{color}
\usepackage{amsfonts,bm,color}
\usepackage{verbatim}
\usepackage{float}
\usepackage{latexsym}
\usepackage{stmaryrd}

\newcommand{\ds}{\displaystyle}
\newcommand{\beq}{\begin{eqnarray}}
\newcommand{\eeq}{\end{eqnarray}}
\newcommand{\beqq}{\begin{eqnarray*}}
\newcommand{\eeqq}{\end{eqnarray*}}

\makeatletter
\newcommand*{\centerfloat}{%
 \parindent \z@
 \leftskip \z@ \@plus 1fil \@minus \textwidth
 \rightskip\leftskip
 \parfillskip \z@skip}
\makeatother

\begin{document}
\newcommand*\samethanks[1][\value{footnote}]{\footnotemark[#1]}
\title{Active unidirectional network flow generates molecular transport in packets}
\author{Matteo Dora and  David Holcman \footnote{\'Ecole Normale Sup\'erieure, 75005 Paris, France}}
\date{}
\maketitle
\begin{abstract}
Internet, social media, neuronal or blood vessel are organized in complex networks. These networks are characterized by several quantities such as the underlying graph connectivity (topology), how they grow in time, scaling laws or by the mean time a random search can cover a network and also by the shortest paths between two nodes. We present here a novel type of network property based on a unidirectional transport mechanism occurring in the Endoplasmic Reticulum network present in the cell cytoplasm. This mechanism is an active-waiting transportation, where molecules have to wait a random time before being transported from one node to the next one. We find that a consequence of this unusual network transportation is that molecules travel together by recurrent packets, which is quite a large deviation behavior compared to classical propagation in graphs. To conclude, this form of transportation is an efficient and robust molecular redistribution inside cells.
\end{abstract}

\begin{figure}
  \centerfloat
  \includegraphics[scale=0.8]{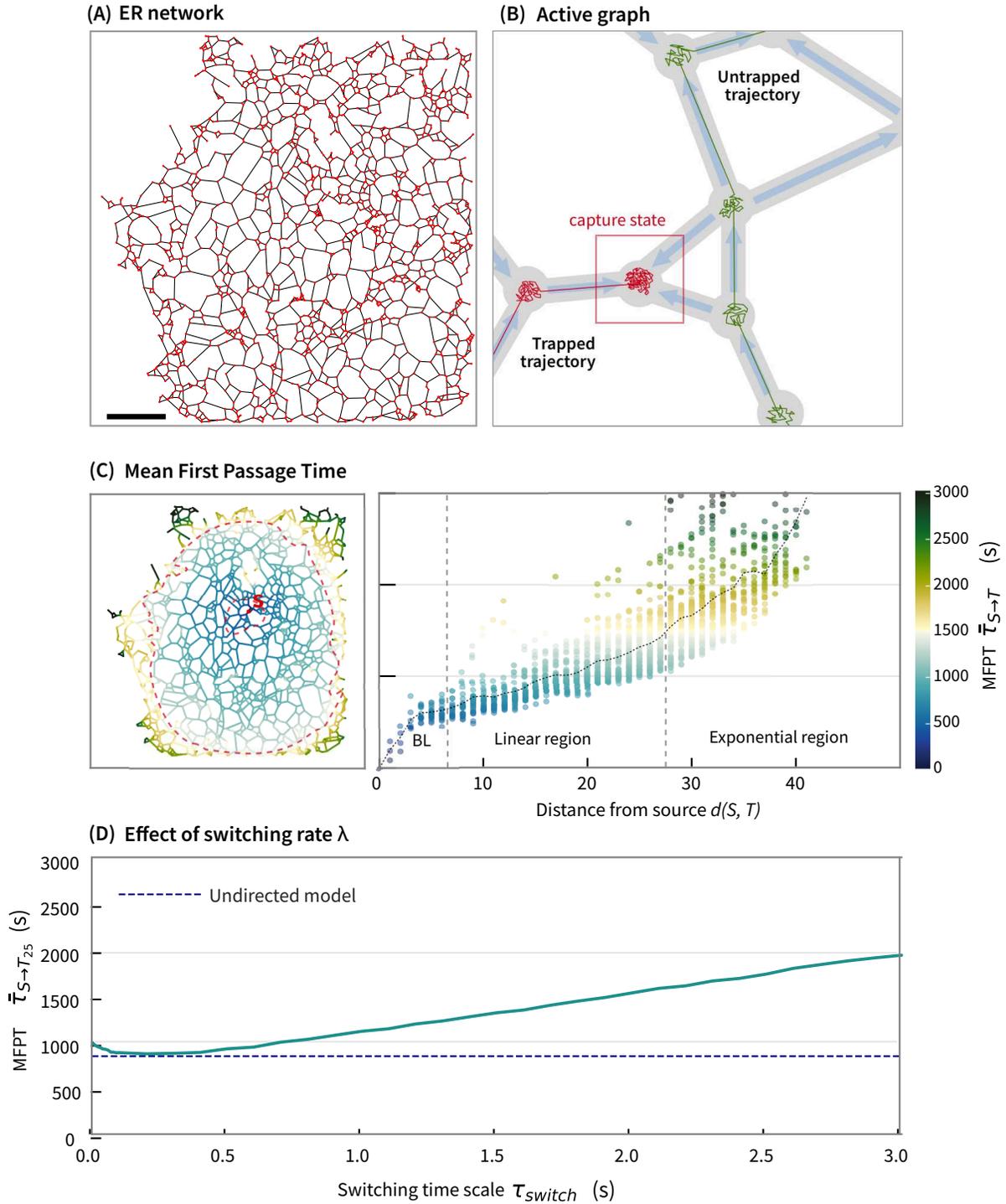}
  \caption{
    \textbf{(A)} ER network reconstructed from SIM data \cite{parutto2018statistical}.
    \textbf{(B)} Active graph model: At a given moment of time, molecular transportation is unidirectional, but this direction switches at random time.
    \textbf{(C)} Heatmap of the mean first passage time (MFPT) (left) and scatter plot of the MFPT for each node ranked by distance from the source (right). Dashed lines separate the boundary layer, linear and exponential regions.
    \textbf{(D)} MFPT vs the switching time $\tau_{switch}$ for a node located at distance 25 from the source. A minimum is observed for a time compatible with the one reported in the ER.
  }
\label{fig:1}
\end{figure}

\noindent Contrary to many graphs such as internet, small-world \cite{amaral2000classes,watts1998collective}, or other complex networks, the Endoplasmic Reticulum (ER), which is a fundamental organelle located inside the cell cytoplasm, consists at steady-state of a network of interconnected tubules which often present three-way junctions  (Fig. \ref{fig:1}a) \cite{terasaki1986microtubules,nixon2016increased}, where each vertex (node or sheet) is connected in average by three edges to three neighboring vertices, with no preferential connectivity. What defines the topology of the ER remains unclear, but the edges are made of small tubules, that could appear and disappear transiently \cite{lee1988dynamic}.
The role of the ER is to redistribute proteins, but it is not clear how this is actually performed and what is the associated time scale \cite{nehls2000dynamics}. Large amount of single particle trajectories data revealed that the direction of the flow in each tubule (network edge) alternates at random time \cite{parutto2018statistical}, leading to an usual redistribution across nodes. Indeed, under this condition, it happens that all the edges incident to a node can have an inward flow, and thus the material in that specific node cannot escape and is trapped until the flow changes direction in at least one of the edges.  We call this temporal trapping situation a capture state (Fig. \ref{fig:1}b). As we shall, transport in this situation is very different from flows or diffusion in classical networks \cite{lovasz1993random}. Although this vertex asymmetry is only transient here, this unidirectional property is reminiscent of the diode networks, introduced in percolation problems by constructing neighbouring lattice sites that transmit connectivity or information in one direction only \cite{broadbent1957percolation,redner1981percolation}. \\
This property makes the ER network quite different from other types of networks and one key parameter is the edge directional switching rate $\lambda$. How protein or molecular trafficking depends on such a rate? A measure of material redistribution is characterized by the mean time it takes for a molecule located in node A to arrive for the first time to node B. We found three regimes (Fig. \ref{fig:1}c): when the node initial node A is close B (boundary layer), the time is then a fast increasing function of the (graph) distance. At intermediate distances, the mean time increases linearly and finally, for regions located far away and connected by very few nodes, the mean time increases drastically. The first region corresponds to $\approx 10$ nodes, while the last one is of the order of the diameter of the graph. Interestingly, the mean first arrival time has a minimum with respect to the switching rate $\lambda$ (Fig. \ref{fig:1}d). \\
To conclude, for a switching rate $\lambda = 30s^{-1}$, it takes about 25 minutes for a molecule to arrive in average to any node. This time scale associated to the mean first arrival time accounts for the total number of nodes of the graph, because on average each trajectory will visit a large portion of the graph before arriving for the first time to the target node, a situation similar to the classical escape through a narrow window \cite{Holcman2015}.
\begin{figure}
\centerfloat
\includegraphics[scale=0.8]{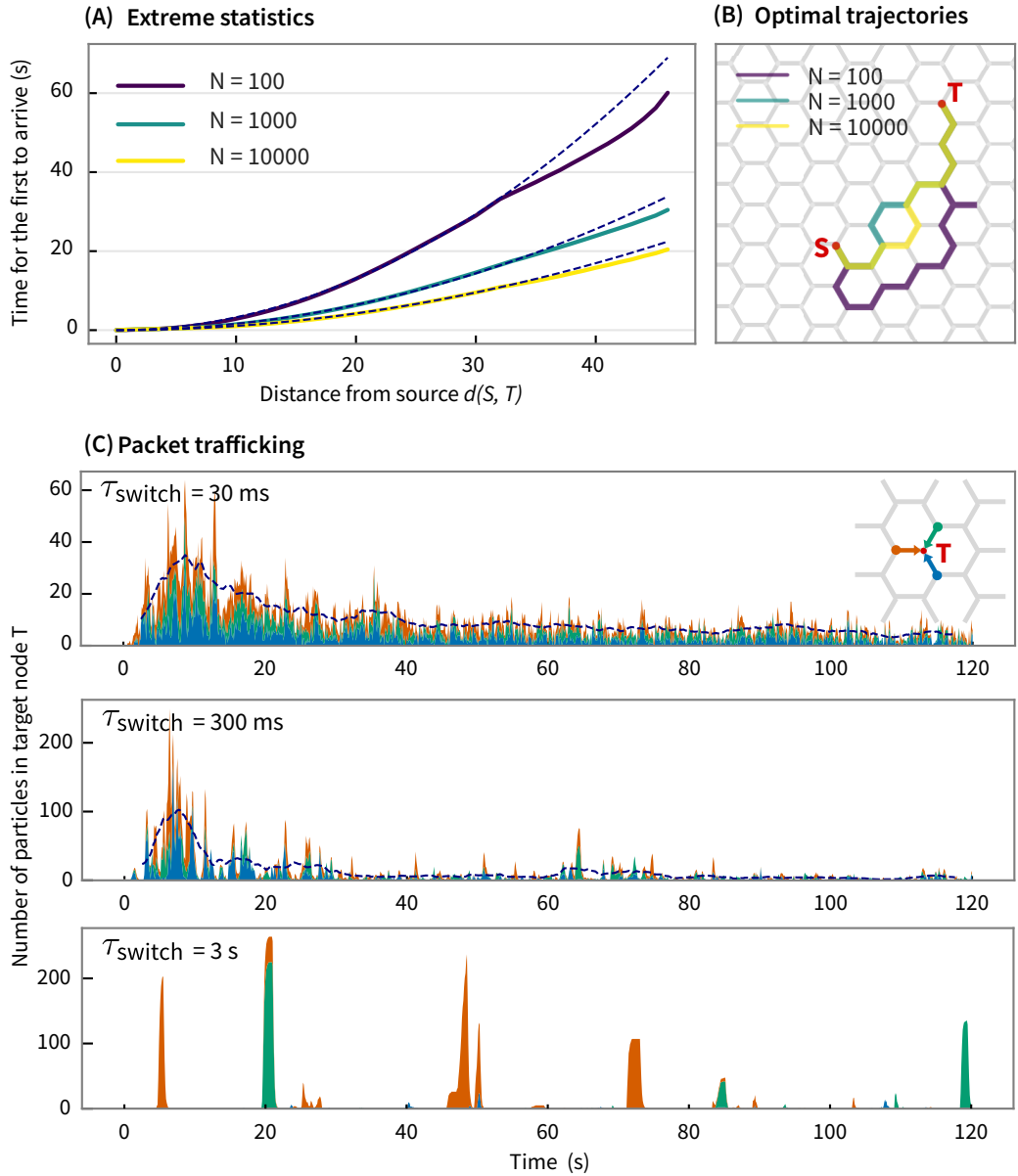}
\caption{ \textbf{(A)} Time for the first particle to arrive at a target node T, when there are initially N particles starting from the source S. The stochastic simulations dashed is compared to fit the law $\frac{c_1 \delta^2}{c_2 + \log N}$,($c_1 = 0.07$ and $c_2 = -2.38$) where $\delta$ is the Graph distance. \textbf{(B)} Gastest trajectories from S to T for different number $N$ of particles released from the source. \textbf{(C)} Number of particles in the target node colored by edge where they are coming from. A classical behavior leading to constant equilibrium distribution is reached for a small switching time scale $\tau_{switch} = 30ms$. However, a new form of trafficking emerges in packets of particles for a time scale of $\tau_{switch} =3s$. This transportation is associated to peaks of density arriving to a node: large groups of particles arrive synchronously from the same edges to a node. This mode of transportation differs significantly from classical transient or steady-state.}
  \label{fig:2}
\end{figure}
Another possible measure of the time scale in such active graph is the arrival time of the fastest particles among many to a given node starting from an initial node. When there is no edge directionality,  particles simply disperse by diffusion with a time scale given by the first eigenvalue (of the Laplace operator on the graph \cite{maier2017cover}). The transient regime follows from the classical diffusion rules, where in the end, the density in all nodes is uniform, equals to the ratio of the number of particles to the total number of nodes $N_{particle}/N_{node}$. However when the edge direction switching is considered, we observed an unexpected phenomena: first, the time scale reduces by two order of magnitude, and does not vary much with the initial number of particles when $N > 1000$. Indeed, the time scale reduces to 20 seconds (Fig. \ref{fig:2}a), compared to 25 minutes for the mean time. The time scale of redistribution is associated to the extreme statistics \cite{majumdar2016exact} leading to the extreme mean time law $\langle \tau\rangle \sim \ds\frac{\delta_{min}^2}{D\log (N_{particle})} $, where $\delta_{min}$ is the graph distance between the source and the target node and $D=\frac{\lambda_{switch} a^2}{3}$ is the effective diffusion coefficient \cite{Holcman2015} (a is the mean distance between two nodes and $\lambda_{switch}$ is the switching rate), as confirmed by stochastic simulations (Fig. \ref{fig:2}a-b). This arrival time for the fastest is quite different from the mean time of arrival for one particle given by $ \langle \tau \rangle \approx \frac{3\sqrt{15} N_{net}}{2\pi \lambda } \ln N_{net}+O(1)$, where the rate $\lambda$ is the reciprocal of the time to switch between two node and $N_{net}$ is the size of hexagonal lattice.\\
Second, due to the capture effect, particles can be trapped many times before reaching the target node. Interestingly, the particles travel by packets that form and deform, arriving to the target node at random times (Fig. \ref{fig:2}c). This situation contrasts with the case of no switching, because the redistribution of material does not converge for long-time regime to the uniform steady-state where all material is shared uniformly by the nodes.\\
To conclude we showed here that the ER dynamics generates an atypical protein redistribution: proteins travel in recurrent packets to any point. This property allows proteins to be delivered in groups, which is probably a more stable delivery process than the arrival of individuals. Moreover the arrival of the fastest particles or packets occurs along the shortest paths (Fig. \ref{fig:2}b), which is the most efficient mode of redistribution. We predicted that the time scale of packets redistribution takes 20-30 seconds. \\
Active networks allow the redistribution of particles by packets, that can arrive to a vertex at different moments of time. This form of redistribution is very different from blood in capillary networks or spikes dispersion in neuronal networks. This mode of delivery, modulated by the switching rate, could have applications to other situations, but already shows that the ER network has evolved to redistribute packets of proteins at a time scale of few seconds, across most of the cytoplasm.\\
Acknowledgements: This research was supported by a FRM grant to D. Holcman. We thank P. Parutto, E. Avezov and D. Ron for discussions.

\normalem
\bibliographystyle{ieeetr}
\bibliography{references2}
\end{document}